\documentclass[12pt]{article}

\usepackage{amsmath}
\usepackage{amssymb}
\usepackage{amsfonts}
\usepackage{dsfont} 

\usepackage{graphicx}

\usepackage[dvipsnames]{xcolor}

\newcommand{\be}{\begin{equation}}
\newcommand{\bel}[1]{\begin{equation}\label{#1}}
\newcommand{\ee}{\end{equation}}
\newcommand{\bea}{\begin{eqnarray}}
\newcommand{\eea}{\end{eqnarray}}
\newcommand{\balign}{\begin{align}}
\newcommand{\ealign}{\end{align}}
\newcommand{\ba}{\begin{array}}
\newcommand{\ea}{\end{array}}
\newcommand{\bfig}{\begin{figure}}
\newcommand{\efig}{\end{figure}}

\newcommand{\eref}[1]{(\ref{#1})}

\newcommand{\bra}[1]{\mbox{$\langle \, {#1}\, |$}}
\newcommand{\ket}[1]{\mbox{$| \, {#1}\, \rangle$}}
\newcommand{\exval}[1]{\mbox{$\langle \, {#1}\, \rangle$}}
\newcommand{\dexval}[1]{\mbox{$\langle \langle \, {#1}\, \rangle\rangle $}}

\newcommand{\rmd}{\mathrm{d}}
\newcommand{\rme}{\mathrm{e}}

\newcommand{\half}{\frac{1}{2}}

\newcommand{\C}{{\mathbb C}}
\newcommand{\R}{{\mathbb R}}

\newcommand{\Z}{{\mathbb Z}}
\newcommand{\N}{{\mathbb N}}

\newcommand{\T}{{\mathbb T}} 

\newcommand{\bfeta}{\boldsymbol{\eta}}

\newtheorem{theo}{Theorem}[section]
\newtheorem{lmm}[theo]{Lemma}
\newtheorem{df}[theo]{Definition}
\newtheorem{prop}[theo]{Proposition}
\newtheorem{cor}[theo]{Corollary}
\newtheorem{rem}[theo]{Remark}
\newcommand{\Proof}{\noindent {\it Proof: }}
\def\qed{\hfill$\Box$\par\medskip\par\relax}

\begin{document}

\title{Universal cumulants and conformal invariance in annihilating random walks with pair deposition}


\author{D Karevski$^{1}$ \and G M Sch\"utz$^{2}$ \and A. Zahra$^{1,2}$}
\vspace{10pt}

\maketitle

\noindent $^{1}$ Laboratoire de Physique et Chimie Th\'eoriques, Universit\'e de Lorraine, CNRS, F-54000 Nancy, France\\
$^{2}$ 
Centro de An\'alise Matem\'atica, Geometria e Sistemas Din\^amicos, Departamento de Matem\'atica, Instituto Superior T'ecnico, Universidade de Lisboa, Av. Rovisco Pais 1, 1049-001 Lisboa, Portugal

\begin{abstract}
We consider annihilating random walks on the finite one-dimensional 
integer torus with deposition of pairs of particles, conditioned on an
atypical jump activity. All cumulants of the activity, defined as the number of particle jumps up to some 
time $t$, are obtained in closed form to leading
order in system size $L$ at the critical point, where in the thermodynamic limit 
the conditioned process undergoes
a phase transition in the universality class of the one-dimensional
quantum Ising model in a transverse field. The generating function
of the cumulants at a distance of order $1/L$ away from the critical point
is proved to be given by two universal quantities, viz., by the central charge 
$c=1/2$ of the Virasoro algebra that characterizes
the Ising universality class and by an explicit universal scaling function.
\end{abstract}


\newpage

\section{Introduction}
\setcounter{equation}{0}
\setcounter{footnote}{0}

Large deviation theory provides a major source of insight into the properties
of physical many-body systems, both in and out of thermal equilibrium. Prominent
examples include dynamical Gibbs-non-Gibbs transitions \cite{vanE10} or fluctuation theorems \cite{Gasp22}
such as the celebrated Evans-Searles Theorem \cite{Evan94}, 
the Jarzinski relation \cite{Evan94,Jarz97}, the Gallavotti-Cohen 
symmetry \cite{Gall95} and violations thereof \cite{Bone06,Harr06}.

In the context of stochastic interacting particle systems (SIPS)
large deviation properties are frequently addressed by conditioning
the Markovian dynamics on some rare event \cite{Feng06,Chet15,Boro17,Mont22}, in particular an atypical value
of the particle current or of the activity, i.e.,
the number of particle jumps
\cite{Lebo99,Derr07,Bert15}. Away from the typical behavior of the 
process, dynamical phase transitions may occur \cite{Bodi05,Garr09,Leco12,Jack15a} that can only be understood 
when the full distribution of current and activity is considered. As noted in 
\cite{Kare17,Kare22}, some such dynamical phase transitions are universal, i.e., 
independent of the microscopic details of the particle system and described 
by conformal field theory, a powerful tool in the study both of static 
equilibrium critical phenomena \cite{DiFr97,Henk99,Card10} and stochastic 
dynamics involving Schramm-Loewner evolution 
\cite{Frie02,Baue03,Card04}. 

Specifically, at higher than typical activity the symmetric simple exclusion 
process (SSEP) \cite{Spit70,Ligg99} defined on an integer torus of $L$ sites can be 
described in the thermodynamic limit $L\to\infty$ by a conformal field 
theory with central charge $c=1$ of the Virasoro algebra \cite{Kare17}
and critical exponents that are given by the Legendre parameter $s$ that is 
conjugate to the activity via Legendre transformation \cite{Mont22}.
At maximal activity this becomes the field theory of free fermions which 
arises also when considering the density fluctuations of a Fermi fluid in 
one dimension which in turn are related to fluctuations of vicinal 
surfaces \cite{Spoh99}. By appealing to universality one thus expects
that independently of microscopic details of the model the same critical 
large-scale behaviour describes generic symmetric lattice gas models with 
one conservation law in the thermodynamic limit $L\to\infty$ of infinite 
system size. 

The universality hypothesis of large deviation properties found support 
some time ago when Appert et al. \cite{Appe08} showed by using the 
Bethe ansatz that the 
cumulants of the particle current and activity distribution of the SSEP 
defined on the integer torus with $L$ sites have universal 
{\it finite-size scaling} properties, i.e., universal large-scale behaviour
when the limit of infinite system size $L$ and Legendre parameter $s- s_c
\propto u/L^z$ is taken simultaneously. For the SSEP, $s_c=0$ is the
phase transition point from a hyperuniform to a phase-separated regime
\cite{Leco12,Jack15a,Kare17} and $z=2$ is a scaling exponent. Also the current distribution in the totally asymmetric simple exclusion process
exhibits universal features with dynamical exponent $z=3/2$ \cite{Arai24,Baik23,Bodi05,Corw12,Mate21,Prae04}. This observation is intriguing and raises the question
what other universality classes (if any) may exist and what universal
scaling function would emerge.

Following this line of research, the study of universality classes for dynamical phase transition in SIPS was further pursued for the process 
of diffusion-limited pair annihilation and deposition (DLPAD).
In this process particles are deposited in pairs in neighboring lattice sites  and perform independent random walks,
but annihilate instantly when two particles meet on the same site.
There is a critical atypical activity, given by a non-zero
Legendre parameter $s_c$ 
at which the process is described again by a conformal field theory \cite{Kare22}, but unlike the conditioned SEP in the universality
class of the critical one-dimensional quantum Ising model in a transverse field with central charge $c=1/2$ \cite{Henk99,Kare00}. 

It is the purpose of the present work to address the question
of universal scaling functions for conditioned processes at a critical
point and their link to conformal invariance rigorously by studying the cumulants of the activity of the DLDAP
at the critical point. 
The main results of present work, established in Theorem \ref{Theo:ucn} and  Theorem \ref{Theo:ucgf}, are an explicit computation of all cumulants
to leading order in system size $L$ and the derivation of a universal generating function $h(u)$ for the cumulants in the thermodynamic limit $L\to\infty$ with distance $s-s_c\propto u/L$ from the critical point $s_c$. The scaling properties are different from their counterparts in the SSEP 
that were obtained in \cite{Appe08}: The scaling function $h(u)$
is different, the scaling exponent 
$z=1$, characteristic for ballistic scaling, is different from the diffusive
exponent $z=2$ for the SSEP, and the odd cumulants scale differently
with system size. 

With a view on future research along these lines we suggest that 
further universal properties and universality classes may be explored mathematically rigorously.
As examples of interest we mention the DLDAP with non-periodic
boundaries, related to the quantum Ising model in a transverse field
with boundary fields \cite{Lieb61,Pfeu70,Kare00}, and the exclusion 
process with nearest neighbour jumps but a logarithmic long-range 
interaction \cite{Spoh99,Popk10} which is closely related to Dyson's 
Brownian motion \cite{Dyso62} and the Calogero-Sutherland model 
\cite{Calo69,Suth72} which in turn all have an intimate well-established 
connection to Coloumb gases and random matrix theory via the circular 
unitary ensemble. Logarithmic interaction potentials play a role also
in stochastic models of DNA denaturation \cite{Bar09,Hirs11}.
Thus understanding universality classes of activity cumulants may shed light
on a large variety of problems of interest in different fields of probability
theory and statistical physics, including applications.

This paper is organized as follows. In Sec. 2 we define the process
in terms of its Markov generator and review its link to the
quantum Ising chain in a transverse field which is used to
prove the results. Also the activity is defined. In Sec. 3 first some
further tools are presented and then the main results 
Theorem \ref{Theo:ucn} and Theorem \ref{Theo:ucgf} as well as 
some special cases of these are stated and proved.

\section{Diffusion-limited pair annihilation and deposition (DLPAD)}
\setcounter{equation}{0}
\setcounter{footnote}{0}

Informally, diffusion-limited pair annihilation and deposition is a 
Markov process where pairs of particles are deposited independently 
on neighboring sites of a lattice $\Lambda$ with a rate $\mu$. These particles perform random walks with rate $w$ to a neighbouring site and 
whenever two particles meet at the 
same site they both annihilate instantly. Hence, at any positive time each site is occupied by at most one particle 
and, as noted in \cite{Beli19}, the DLPAD starting with at most one particle
per site
is equivalent to a nonconservative exclusion process where 
(i) particles jump with rate $w+\mu$
 to nearest neighbour sites provided the target site
is empty, 
(ii) neighbouring pairs of particles are annihilated with rate $2w +\mu$,
(iii) particle pairs are created with rate $\mu$, provided the pair of 
target sites is empty.

This nonconservative exclusion process is a SIPS
that was first studied in some detail in \cite{Gryn94}.
For $\mu=0$  this process reduces to diffusion-limited pair annihilation,
also known as annihilating random walk, and for $w=0$ to pair annihilation and deposition which
both have a long history of study, see e.g. \cite{Sudb95,Lloy96,Schu01}
and references therein. In the present work both $w$ and $\mu$ are 
taken to
be  strictly positive. As lattice we consider the 
finite integer torus $\T_L = \Z/(L\Z)$ with $L \geq 2$ sites
and assume that the process starts with at most one particle per site.

\subsection{Generator and invariant measure of the DLPAD}
\label{Sec:DefDLPAD}

The lattice sites $k\in \T_L$ of the finite integer torus $\T_L = \Z/(L\Z)$ with $L \geq 2$ sites are counted from $0$ to 
$L-1$ modulo $L$. Each 
may be empty or occupied by at most one particle which is indicated by the
occupation number $\eta_{k} \in \{0,1\}$. A complete configuration of the lattice
is specified by the set of all occupation numbers $\bfeta := (\eta_{0}, \dots, 
\eta_{L-1}) \in \{0,1\}^{L}$. The locally flipped configuration $\bar{\bfeta}^{ll+1}$ is defined for a given 
configuration $\bfeta$ by the occupation numbers 
\begin{eqnarray} 
\bar{\eta}^{ll+1}_{k} & = & \left\{ \ba{ll} 1-\eta_{k} & \mbox{if } k \in \{l,l+1\} \\[2mm] \eta_{k} & \mbox{else.} \ea \right.
\end{eqnarray}
and is used to define transitions from one configuration to another.

\begin{df} 
With the local transition rate
\begin{equation}\label{rateDLPAD}
w_{k}(\bfeta) := \mu + w \left(\eta_{k} + \eta_{k+1}\right) \in \R^+.
\end{equation}
with strictly positive jump rate $w$ and strictly positive deposition rate 
$\mu$
the Markov chain on the state space $\Omega_L:\{0,1\}^L$ defined by the generator
\begin{equation}\label{generatorDLPAD}
(\mathcal{H} f) (\bfeta) = \sum_{k=0}^{L-1} w_{k}(\bfeta)
\left[f(\bar{\bfeta}^{kk+1}) - f(\bfeta)\right]
\end{equation}
acting on bounded functions $f:\Omega_L\to\R$ 
is called diffusion-limited pair annihilation and deposition (DLPAD).
\end{df}

\begin{rem}
Since particles are annihilated or deposited in pairs, the total particle number
\begin{equation}\label{DLPADpartnum}
N(\bfeta) = \sum_{k=0}^{L-1} \eta_{k}
\end{equation}
is even or odd at all positive times. Hence the DLPAD is trivially nonergodic, with disjoint ergodic sectors with an even (+) or
odd (-) number of particles respectively. We denote the respective
state spaces by $\Omega_L^\pm$ and the 
generators restricted to these state spaces by $\mathcal{H}^\pm$.
\end{rem}

It was shown in  \cite{Gryn94} that the DLPAD is 
reversible w.r.t. the Bernoulli product measure
\begin{equation}\label{DLPADstat}
\pi(\bfeta) = \prod_{k=1}^L b(\eta_{k})
\end{equation}
with the marginals
\begin{equation}\label{DLPADstatsingle}
b(\eta) = 
(1-\eta)(1-\rho) + \eta \rho
\end{equation}
with particle density
\begin{equation}\label{statdens}
\rho 
= \frac{1}{1+ \sqrt{1+\frac{2w}{\mu}} }.
\end{equation}
For any non-zero deposition rate $\mu$ the density is in the
range $0<\rho<1/2$.

\subsection{Intensity matrix and quantum XY chain}

As detailed in \cite{Lloy96,Schu01,Schu19}, the intensity matrix (frequently called $Q$-matrix
in the probabilistic literature \cite{Ligg10}) defined by the generator of DLPAD has a natural
representation in terms of the Hamiltonian operator of a non-Hermitian
quantum spin chain. According to \cite{Ligg10} the
off-diagonal matrix elements $Q_{\bfeta \bfeta'}$ of the intensity matrix $Q$ are the transition 
rates 
\begin{equation}
\label{transitionrate}
Q_{\bfeta \bfeta'} = \sum_{l=0}^{L-1} w_l(\bfeta) \delta_{\bfeta',\bar{\bfeta}^{ll+1}}
\end{equation}
from a configuration $\bfeta$ to configuration $\bfeta'$ of the DLPAD
with the local transition rates $w_l(\bfeta)$ defined in \eqref{rateDLPAD}.
The diagonal elements are given by
\be
Q_{\bfeta\bfeta} = - \sum_{\bfeta' \in \Omega_L\setminus\bfeta} w_{\bfeta' \bfeta} 
\end{equation}
which expresses conservation of probability.
The matrix $H= - Q^T$ with offdiagonal
elements $H_{\bfeta'\bfeta} = - Q_{\bfeta\bfeta'}$ is a
non-Hermitian matrix related to the Hamiltonian operator of the
one-dimensional spin-1/2 quantum XY-chain in a transverse field 
\cite{Gryn95} given by 
\begin{equation}\label{HDLPAD}
H =  \sum_{k=0}^{L-1} h_{k,k+1}
\end{equation}
with 
\begin{eqnarray}
h_{k,k+1} & = & 
- (w+\mu) \left(\sigma^+_{k} \sigma^-_{k+1} + \sigma^-_{k} \sigma^+_{k+1}\right) 
\nonumber \\ & & 
- (2w+\mu) \sigma^+_{k} \sigma^+_{k+1} - \mu \sigma^-_{k} \sigma^-_{k+1} \nonumber \\
& & + w(\hat{n}_{k} + \hat{n}_{k+1})
+ \mu \mathbf{1}
\label{Hlocal}
\end{eqnarray}
expressed in terms of the unit matrix $\mathbf{1}$
and the spin-1/2 ladder operators 
$\sigma^\pm_k = (\sigma^x_k \pm i \sigma^y_k)/2$ and projectors
$\hat{n}_{k} = (\mathbf{1} - \sigma^z_k)/2$ acting on site $k$, see \cite{Lloy96,Schu19}
for a review of the representation of the generators of stochastic interacting particle
systems in terms of quantum spin chains and specifically \cite{Beli19} for a detailed
derivation of \eref{HDLPAD} from the generator \eref{generatorDLPAD}.
Here we only mention that in this construction one uses the canonical basis vectors 
$\mathfrak{e}_{0} = (1,0)^T$ and $\mathfrak{e}_{1} = (0,1)^T$ of $\C^2$
from which one constructs the tensor basis 
\begin{equation}\label{tensorbasis}
\ket{\bfeta} := \mathfrak{e}_{\eta(0)}
\otimes \mathfrak{e}_{\eta(1)} \otimes \dots \mathfrak{e}_{\eta(L-1)}
\end{equation}
of $(\C^{2})^{\otimes L}$ which is isomorphic to $\C^{2^L}$
and which provides an isomorphism between the elements $\bfeta$
of the state space $\Omega_L$ and the canonical basis vectors 
$\ket{\bfeta}$ and their transposed vectors $\bra{\bfeta} := \ket{\bfeta}^T$ of $\C^{2^L}$.
In slight abuse of language we call $H$ the quantum
Hamiltonian of the process.

The projections on the even and odd particle sectors can be constructed
in terms of the particle numebr operator $\hat{N}$, particle parity operator $\hat{P}$, and the projectors
$\hat{P}^\pm$ defined by
\begin{equation}
\hat{N} := \sum_{k=0}^{L-1} \hat{n}_k, \quad 
\hat{P} := (-\mathbf{1})^{\hat{N}}, \quad 
\hat{P}^{\pm} := \half \left(\mathbf{1} \pm \hat{P}^{\pm}\right).
\end{equation}
Then
\begin{equation}
H^{\pm} := H \hat{P}^{\pm}
\end{equation}
yield the intensity matrices $Q^{\pm} = - (H^{\pm})^T$ that define
the Markov semigroup associated with the DLDAP in the even ($+$) and 
odd ($-$) sector respectively.

%
The link to the Hermitian quantum Hamiltonian operator
of the one-dimensional spin-1/2 quantum XY-chain in a transverse field 
enables us to use various well-known results \cite{Henk99}. 
To this end, it is useful to introduce the spin operator 
$\hat{S}^z$ defined by
\begin{equation}\label{Szdef}
\hat{S}^z :=  \sum_{k=0}^{L-1} \sigma^z_k = L  - 2 \hat{N} 
\end{equation}
and the parameters
\begin{equation}\label{etahdef}
J:= w+\mu, \quad \gamma := 
\sqrt{\frac{\mu (2w+\mu)}{(w+\mu)^2}}
, \quad
h := 
\frac{w}{w+\mu}, \quad 
z:= 
\sqrt{\frac{\mu}{2w+\mu}}.
\end{equation}

With the diagonal matrix
\begin{equation}\label{piastmatrix} 
\hat{\pi} :=  z^{\hat{N}}
\end{equation}
one obtains
\begin{eqnarray}
H^{XY} & := & \hat{\pi}^{-\half} H \hat{\pi}^{\half}
\label{tHdef} \\
& = &  -  J \sum_{k=0}^{L-1} 
\left(  \frac{1 + \gamma}{2} \sigma^x_{k} \sigma^x_{k+1} 
+  \frac{1 - \gamma}{2} \sigma^y_{k} \sigma^y_{k+1} 
+ h \sigma^z_k - 1\right)
\label{tHDLPAD} 
\end{eqnarray}
which is the 
quantum Hamiltonian operator
of the one-dimensional spin-1/2 quantum XY-chain in a transverse field
\cite{Lieb61,Niem67,Pfeu70,Henk87}.


%

\subsection{Activity}

\subsubsection{Definition}
The activity $A(t)$ is defined as the total number of particle jumps in
the DLPAD that have occured up to time $t>0$, starting from some
initial configuration with at most one particle on each site. Hence
$A(t)\in\N_0$ is a random number that is incremented by 1 whenever
a particle jump with rate $w$ occurs and $A(0)=0$. Following the
strategy outlined in \cite{Harr07}, the evolution
of $A(t)$ can be captured by
the joint process $(\bfeta(t),A(t))$ with state space
$\tilde{\Omega}_L := \Omega_L \times \N_0$ defined by the generator
\begin{eqnarray}
(\tilde{\mathcal{H}} f)(\bfeta,A) & = & \sum_{k=0}^{L-1} 
\left[\mu f(\bar{\bfeta}^{kk+1},A) + w(\eta_{k} + \eta_{k+1}) 
f(\bar{\bfeta}^{kk+1},A+1) 
 \right. \nonumber \\
& & \left. 
- [\mu+ w(\eta_{k} + \eta_{k+1})] f(\bfeta,A)\right]
\label{generatorADLPAD}
\end{eqnarray}
for all $\bfeta\in\Omega_L$ and $A\in\N_0$.
Denoting expectations for this joint process with initial configuration
$\bfeta$ by $ \dexval{\cdot}_{\bfeta,0}$, the moment generating function of the
activity is defined by
\begin{equation}
G_{L}(\bfeta,s,t) := 
\dexval{\rme^{s A(t)}}_{\bfeta,0} 
\end{equation}
where $s$ is the Legendre parameter conjugate to the activity.
Expressing the activity distribution in terms of the Legendre parameter
is analogous to passing from a microcanonical ensemble for fixed $A$
to a canonical fluctuating ensemble \cite{Mont22}.

As established in \cite{Chet15} this generating function
can be expressed in terms of a {\it tilted generator} which for the
DLPAD reads
\begin{equation}
(\mathcal{H}_{s} f)(\bfeta) = \sum_{k=0}^{L-1} 
\left[(\mu + w \rme^{s} (\eta_{k} + \eta_{k+1}))  f(\bar{\bfeta}^{kk+1})
- w_k(\bfeta) f(\bfeta,A)\right]
\label{tgeneratorDLPAD}
\end{equation}
acting on bounded functions defined on the original state space
$\Omega_L$. One has 
\begin{equation}
G_{L}(\bfeta,s,t) = \sum_{\bfeta' \in \Omega_L}\exval{\mathbf{1}_{\bfeta'}(t)}_{\bfeta}
\end{equation}
where the expectation $\exval{\cdot}_{\bfeta}$ is taken for the evolution defined by the
tilted generator $\mathcal{H}_{s}$ which does {\it not}
conserve probability so that the expectation over the sum of all
indicators is not equal to 1 but a non-trivial
function of $s$ and $t$. 

\subsubsection{Tilted intensity matrix}

Associated with the tilted generator is the tilted intensity matrix 
$Q(s)$ and the tilted quantum Hamiltonian
\begin{eqnarray}
H_{s} 
 & = & - Q^T_s \, = \, - \sum_{k=0}^{L-1} \left[
(w \rme^{s} + \mu)  \left(\sigma^+_{k} \sigma^-_{k+1} + \sigma^-_{k} 
\sigma^+_{k+1}\right)  
\right. \nonumber \\
 & &  \left.
+  (2 w \rme^{s} +\mu) \sigma^+_{k} \sigma^+_{k+1} 
 + \mu \sigma^-_{k} \sigma^-_{k+1} 
+ w \sigma^z_k - (w+\mu)\right] 
\label{Gtiltact}
\end{eqnarray}
which is of the same form as the untilted quantum Hamiltonian
\eqref{HDLPAD}
but with $s$-dependent parameters. Replacing in the diagonal transformation matrix \eqref{piastmatrix} the parameter $z$
by 
\begin{equation}
z_{s} := \sqrt{\frac{\mu}{2 w \rme^{s} +\mu}}
\end{equation}
therefore yields the transformed quantum Hamiltonian
\begin{equation}
H^{XY}_{s} 
= - J_{s} \sum_{k=0}^{L-1} 
\left(  \frac{1 +  \gamma_{s}}{2} \sigma^x_{k} \sigma^x_{k+1} 
+  \frac{1 - \gamma_{s}}{2} \sigma^y_{k} \sigma^y_{k+1} 
+ h_{s} \sigma^z_k - 1\right)
\label{HXY}
\end{equation}
where
\begin{equation}
J_{s} := w \rme^{s} + \mu , \quad
\gamma_{s} :=  \sqrt{\frac{\mu  (2 w \rme^{s} +\mu)}{(w \rme^{s} +\mu)^2}}, \quad 
h_{s} := \frac{w}{w \rme^{s}+ \mu}
\label{psdef1}
\end{equation}
and $H^{XY}_{0} = H^{XY} $.

In terms of the tilted quantum Hamiltonian the moment generating 
function is given by 
\begin{equation}
G_{L}(\bfeta,s,t) = \sum_{\bfeta' \in \Omega_L}
\bra{\bfeta'} \rme^{-H^{XY}_s t} \ket{\bfeta}
\end{equation}
which follows from the discussion in \cite{Chet15} by using the
matrix representation of the generator in the tensor basis \eqref{tensorbasis} defined above.
For $\mu < 1$ the ground state of this Hamiltonian, discussed below, has a critical point
\begin{equation}
s_c := \ln{\left(1 - \mu/w\right)} < 0
\label{scdef}
\end{equation}
at a finite negative value of $s$ where
the Ising phase transition referred to in the introduction occurs. 


\subsubsection{Cumulant generating function}

As $A(t)$ is extensive both in system length $L$ in time $t$ we consider the scaled activity $a(t):=A(t)/(Lt)$ with 
moment generating function $g_{L}(\bfeta,s,t) := 
\dexval{\rme^{s a(t)}}_{\bfeta,0}$ and cumulant generating
function $K_{\bfeta}(L,s,t) := \ln{g_{L}(\bfeta,s,t)}$. 
As in \cite{Appe08} we are
interested in the asymptotic long-time behavior $t\to\infty$
at fixed $L$ and $s$
where due to ergodicity the moment generating function 
and therefore also the cumulant generating function 
lose their dependence on the initial configuration except for the
initial particle number parity. Thus we define for $\bfeta\in\Omega_L^{\pm}$
\begin{equation}
K^\pm_{L}(s) := \lim_{t\to\infty} \left.K_{\bfeta}(L,s,t)\right|_{\bfeta\in\Omega_L^{\pm}}
\end{equation}
and note the trivial property $K^\pm_{L}(s) = 0$ for $\bfeta\notin\Omega_L^{\pm}$.

The second point to observe is that since the state space is finite
the tilted quantum Hamiltonian has a finite spectral gap in each sector
defined by the projected quantum Hamiltonian $H_{s}^{\pm} 
:= H_{s} \hat{P}^{\pm}$
with lowest
eigenvalues denoted by $E^\pm_{L}(s)$. Since $E^\pm_{L}(s)$ is extensive in
system size $L$ it is natural to introduce the ground state energy
per site $\epsilon^\pm_{L}(s) = E^\pm_{L}(s)/L$. Then the limiting
cumulant generating function of $a(t)$ is in complete analogy to \cite{Appe08}
given by
\begin{equation}
K^\pm_{L}(s) = - \epsilon^\pm_{L}(s).
\end{equation}

The ground state energy per site has been computed exactly using
techniques based on the Jordan-Wigner transformation
that expresses the spin lowering and raising operators appearing
the quantum Hamiltonian into operators satisfying fermionic anticommutation
relations \cite{Lieb61}. This transformation leads to a 
many-body system
of spinless free fermions that can be solved exactly
using the techniques developed in \cite{Lieb61} or, more straightforwardly, by the Cooper pair approach of \cite{Schu95}
that by duality reduces the computation of the ground state energy 
of the many-body DLPAD to treating a collection of non-interacting two-state 
systems \cite{Schu20}.
With the discrete dispersion relation \cite{Niem67,Kare22}
\begin{eqnarray}
\Lambda^{+}_{L,r}(s) & = &  J_{s} 
\sqrt{\left[h_{s} - \cos{\frac{\pi (2r+1)}{L}}\right]^2 + \left[\gamma_{s} \sin{\frac{\pi (2r+1)}{L}}\right]^2 } \\
\Lambda^{-}_{L,r}(s) & = & J_{s} 
\sqrt{\left(h_{s} - \cos{\frac{2 \pi r}{L}}\right)^2 + \left(\gamma_{s} \sin{\frac{2 \pi r}{L}}\right)^2 } , \quad r \neq 0 \\
\Lambda^{-}_{L,0}(s) & = & J_s
\label{betahr}
\end{eqnarray}
which satisfies the symmetry relation $\Lambda^\pm_{L,r}(s) = \Lambda^\pm_{L,L-1-r}(s)$
one gets \cite{Kare00,Kare22}
\begin{equation}
\label{gse}
K^\pm_{L}(s) = - (w+\mu) 
+ \frac{2}{L}  \sum_{r=0}^{\frac{L}{2}-1}  \Lambda^\pm_{L,r}(s).
\end{equation}
The cumulants are then given by
\begin{equation}
\kappa^\pm_{L,n}(s) :=  \frac{d^n}{ds^n}K^\pm_{L}(s)
= \left\{\ba{ll} \displaystyle
\frac{2}{L}  \sum_{r=0}^{\frac{L}{2}-1} \frac{d^n}{ds^n} \Lambda^{+}_{L,r}(s) & \mbox{even sector} \\
\displaystyle \frac{2w \rme^{s}}{L}  + \frac{2}{L}  \sum_{r=1}^{\frac{L}{2}-1} \frac{d^n}{ds^n} \Lambda^{+}_{L,r}(s) & \mbox{odd sector}
\ea\right.
\label{kpmnLs}
\end{equation}
We recall that the first cumulant $K^\pm_1(L,s)$ is the mean activity, 
the second cumulant is the variance, and the third
cumulant is the third central moment.

Below we focus on the even sector at the critical point $s=s_c$
defined in \eqref{scdef} and
to ease notation the superscript $+$ is dropped on all quantities
considered below. 
In terms of the rates of the
DLPAD and with the short-hand notation
\begin{eqnarray}
\mathsf{c}_{L,r} := \cos{\frac{\pi(2r + 1)}{2L}}, \quad 
\mathsf{s}_{L,r} := \sin{\frac{\pi(2r + 1)}{2L}}
\label{cspdef} 
\end{eqnarray}
we note that in terms of the rates $w$ and $\mu$ of the DLDAP
\begin{equation}
\Lambda_{L,r}(s) 
= \sqrt{\left[ w
- (w \rme^{s} +\mu) \mathsf{c}_{L/2,r}\right]^2 +  (2 w \rme^{s} +\mu) \mu
(\mathsf{s}_{L/2,r})^2} .
\label{disp}
\end{equation}
Since for $w=0$ the activity is trivially
zero only $w>0$ needs to be considered. We introduce the
normalized deposition rate $\tilde{\mu} = \mu/w$ which
makes $w$ the overall time scale of the process. Without loss
of generality we take $w=1/2$ which means that particles
jump after an exponential random time with parameter $1$.
It is convenient to express the results for the critical point in terms of 
the complementary deposition rate
\begin{equation}
\nu =  1- \tilde{\mu}
\label{nudef}
\end{equation}
that appears frequently in the explicit formulas obtained below.
We consider $\nu$ to be fixed and therefore omit the
dependence of the cumulants and other functions on this quantity.

After these preparations we can state and prove the main results,
all of which are based on taking judiciously chosen limits
in \eqref{gse}. The proofs are all computational.

\section{Critical cumulants}

The general expression \eqref{gse} yields
the cumulant generating function 
\begin{equation}
K_L^{c}(x) := K_L(s_c+x) - K_L(s_c) = \sum_{n=1}^\infty \kappa^{c}_{L,n} \frac{x^n}{n!}
\label{KcLdef}
\end{equation}
for the cumulants 
\begin{equation}
\kappa^{c}_{n}(L) := \kappa_{L,n}(s_c)
\label{kcnLdef}
\end{equation}
at the critical point \eqref{scdef}. To highlight the anticipated significance
of the large-$L$ behaviour of the cumulants at the critical point
the dependence on $L$ is written as an argument rather than a
subscript.

\subsection{Auxiliary Lemmas}

For large system size,
the leading order behaviour of the cumulant generating function 
evaluated at the critical point is captured by using the following Lemma.
\begin{lmm} 
\label{Lem:BmN}
To the first subleading order in $1/N$
the trigonometric sums
\begin{equation}
\Phi_m(N) := \frac{2}{N} \sum_{k=0}^{N-1} \sin\left(\frac{\pi (2 k+1)}{4N}\right)
\cos^{2m}\left(\frac{\pi (2 k+1)}{4N}\right)
\label{BpmNdef} 
\end{equation}
are for $m\geq 0$ and $N\geq 1$ given by
\begin{equation}
\Phi_m(N)
= \frac{4}{\pi (2m+1)}  +  \frac{\pi }{24 N^2} + O(N^{-4}).
\label{BpmN} .
\end{equation}
\end{lmm}

\Proof By definition and binomial theorem
\begin{eqnarray*}
\Phi_m(N) 
& = & \frac{1}{N 2^{2m} i} \sum_{k=0}^{N-1}
\rme^{i\frac{\pi (2 k+1)(2m+1)}{4N}} 
\left(1 - \rme^{-2i\frac{\pi (2 k+1)}{4N}}\right)
\left(1 + \rme^{-2 i\frac{\pi (2 k+1)}{4N}}\right)^{2m} \\
& = & \frac{1}{N 2^{2m} i} \sum_{r=0}^{2m} {2m \choose r} \sum_{k=0}^{N-1}
\rme^{i\frac{\pi (2 k+1)(2m+1-2r)}{4N}} 
\left(1 - \rme^{-2i\frac{\pi (2 k+1)}{4N}}\right)
\end{eqnarray*}
Using the geometric sum formula yields for $m\geq 0 $ and any finite $N\geq 1$
\begin{eqnarray}
\Phi_m(N) 
& = & \frac{1}{N 2^{2m} i} \sum_{r=0}^{2m} {2m \choose r} \left( \rme^{i\frac{\pi (2m+1-2r)}{4N}} \frac{1-\rme^{i\frac{\pi (2m+1-2r)}{2}}  }{1-\rme^{i\frac{\pi (2m+1-2r)}{2N}}  }  \right. 
\nonumber \\
& & \left.
- \rme^{i\frac{\pi (2m-1-2r)}{4N}} \frac{1-\rme^{i\frac{\pi (2m-1-2r)}{2}}}{1-\rme^{i\frac{\pi (2m-1-2r)}{2N}}}
 \right) \nonumber \\
& = & \frac{-1}{N 2^{2m+1}} \sum_{r=0}^{2m} {2m \choose r} \left(  \frac{1-i (-1)^{m-r} }{ \sin{\frac{\pi (2r-2m-1)}{4N}} }
- \frac{1+i (-1)^{m-r}}{\sin{\frac{\pi (2r-2m+1)}{4N}} }
 \right) \nonumber \\
& = & \frac{1}{N 2^{2m+1}} \sum_{r=0}^{2m} {2m \choose r} \left(  \frac{1-i (-1)^{m-r} }{ \sin{\frac{\pi (2m+1-2r)}{4N}} }
+ \frac{1+i (-1)^{m-r}}{\sin{\frac{\pi (2m+1-2r)}{4N}} }
 \right) \nonumber \\
& = & \frac{1}{N 2^{2m}} \sum_{r=0}^{2m} {2m \choose r}  \frac{1}{ \sin{\frac{\pi (2m+1-2r)}{4N}} } .
\label{BpmNexact}
\end{eqnarray}
The Taylor expansion
\begin{equation}
\frac{1}{\sin{x}} = \frac{1}{x} + \frac{x}{6} + O(x^{3})
\end{equation}
and observing that
\begin{equation*}
\sum_{r=0}^{2m} {2m \choose r} \frac{1}{ 2r-2m+1}
= \sum_{r=0}^{2m} {2m \choose r} \frac{1}{2m+1-2r}
\end{equation*}
then yields the two leading order terms
\begin{eqnarray*}
\Phi_m(N) 
& = & \frac{1}{2^{2m-1}\pi } \sum_{r=0}^{2m} {2m \choose r} 
\left(\frac{1}{2m+1-2r} - \frac{1}{2m-1-2r}\right) \\
& & + \frac{1}{2^{2m+1}} \sum_{r=0}^{2m} {2m \choose r} 
  \frac{\pi }{12N^2} \\
& = & \frac{1}{2^{2m-1} \pi}  \left(
\sum_{r=1}^{2m} \frac{{2m \choose r} -  {2m \choose r-1}}{2m+1-2r}
+ 2 \right) +  \frac{\pi }{24N^2} \\
& = & \frac{1}{2^{2m-1} \pi (2m+1)}  \left( 
\sum_{r=1}^{2m} {2m+1\choose r} + 2
 \right) +  \frac{\pi }{24N^2} 
\end{eqnarray*}
Since the sum over the binomial coefficients adds up to $2^{2m+1}$,
the assertion \eqref{BpmN} follows.
\qed

For an exact computation of the cumulants we use 
Faà di Bruno's Theorem \cite{Comt74}, which is a generalisation of the chain rule to arbitrary order, viz.,
\begin{equation}
\frac{d^n}{dx^n} g(f(x)) = \sum_{k=1}^n \left.\frac{d^k}{dy^k} g(y)\right\vert_{y=f(x)} \cdot B_{n,k}\left( f'(x), f''(x), \ldots, f^{(n-k+1)}(x) \right)
\end{equation}
with the Bell polynomials 
\begin{equation}
B_{n,k}(x_1, x_2, \dots, x_{n-k+1}) 
= n! \sum_{j_1, j_2, \dots, j_{n-k+1}} \prod_{i=1}^{n-k+1}
\frac{x_i^{j_i}}{(i!)^{j_i} (j_i)!}
\label{Belldef1}
\end{equation}
where
the sum is taken over all nonnegative integers $j_1, j_2, \dots, j_{n-k+1}$ such that 
\begin{equation}
j_1 + j_2 + \cdots + j_{n-k+1} = k \quad \text{and} \quad j_1 + 2j_2 + \cdots + (n-k+1)j_{n-k+1} = n.
\label{Belldef2}
\end{equation}
The cases relevant are the derivatives of a function of the exponential function and the square root of a quadratic function 
from which the cumulant generating function \eqref{KcLdef} 
and the cumulants \eqref{kcnLdef} can be 
computed for any value of system size $L$ and of the expansion parameter $x$. 

\begin{lmm}
\label{Lemma:FdB1}
Let $g(\cdot)$ be $n$ times differentiable. Then
\begin{equation}
\frac{d^n}{dx^n}  g(\rme^x) =   \sum_{m=1}^{n} S(n,m) \rme^{mx} \left.\frac{d^m}{dy^m}  g(y)\right\vert_{y=\rme^{x}}
\end{equation}
where 
\begin{equation}
S(n,m) =  \frac {1}{m!} \sum _{j=0}^{m}(-1)^{m-j}{\binom{m}{j}} j^{n} 
\label{FdB1}
\end{equation} 
is the Stirling number of the second kind. 
\end{lmm}

\Proof This follows from  Faà di Bruno's formula by using the
combinatorial  identity $S(n,m) = B_{n,m}(1,\dots,1)$ and
the first constraint  \eqref{Belldef2} in the definition of the Bell polynomials.
\qed

\begin{lmm}
\label{Lemma:FdB2}
Let $f(\cdot)$ be a quadratic polynomial with roots $x_0,x_1$. Then for $n\geq 1$ and $x\notin\{x_0,x_1\}$
\begin{eqnarray}
\frac{d^n}{dx^n} \sqrt{f(x)} 
& = & 
\sum_{k= \lfloor \frac{n+1}{2} \rfloor}^n
 \frac{(-1)^{k+1}}{2^{n+k-1}} 
\frac{n!(2k-2)!}{(2k-n)! (n-k)!(k-1)!} \nonumber \\
& & \times
\frac{\left(f'(x)\right)^{2k-n} \left( f''(x)\right)^{n-k}}{\left(f(x)\right)^{k-1/2}}
\label{FdB2}
\end{eqnarray}
where $\lfloor . \rfloor$ is the integer part function.
\end{lmm}


\Proof 
Since by assumption $f(\cdot)$ is a polynomial of order 2, only the Bell polynomials
$B_{n,k}(x_1, x_2,0 \dots, 0)$ 
for which the summation is constrained to
\begin{equation}
j_1 + j_2 = k, \quad j_1 + 2j_2 = n, \quad j_3=\dots=j_{n-k+1}=0
\end{equation}
appear in Faà di Bruno's formula. Therefore $j_2 = n - k$, $j_1 =  2k -n$
which yields
\begin{equation}
B_{n,k}(x_1, x_2,0, \dots, 0) = \frac{n!}{(2k -n)!(n-k)! } x_1^{2k-n} 
\left( \frac{x_2}{2} \right)^{n-k}, \quad 
\end{equation}
in the range $0 \leq k \leq n \leq 2k$ or, equivalently,
\begin{equation}
0 \leq \frac{n}{2} \leq k \leq n
\end{equation}
which is the summation range in \eqref{FdB2}.
Since
\begin{equation}
    \frac{d^k}{dy^k} \sqrt{y} = (-1)^{k+1} \frac{(2k-2)!}{(k-1)!} \left(\frac{1}{2\sqrt{y}}\right)^{2k-1}
\end{equation}
the assertion \eqref{FdB2} follows.
\qed

\subsection{Cumulant generating function at criticality}

The cumulant generating function $K_{L}(s_c)$ at the critical point is given by the sum \eqref{gse}
where at the critical point the dispersion relation \eqref{disp} can be written
as
\begin{eqnarray}
\lambda^{(0)}_{L,r} := \Lambda_{L,r}(s_c) 
& = & \mathsf{s}_{L,r}
\sqrt{1 -  \nu^2 \mathsf{c}_{L,r} } 
\label{l0def}
\end{eqnarray}
 in terms of the complementary deposition rate $\nu$ defined in 
\eqref{nudef}.
Here the trigonometric identities
$\sin{2p} = 2 \sin{p} \cos{p}$ and $1-\cos{2p} = 2 \sin^2{p}$
are used and it is also taken into account that in the summation range
$0 \leq k \leq L/2 - 1$ the sine function does not change sign. 

\begin{prop}
\label{Prop:gsed}
To leading and subleading orders in system size $L$
the critical cumulant generating function in the range $0\leq \nu \leq 1$ 
take the value
\begin{eqnarray}
K_{L}(s_c) & = & \kappa^\ast_0 +  \frac{1}{\xi}\frac{\pi c}{6 L^2}
+ O(L^{-4})
\label{gsep} \\
K^{-}_{L}(s_c) & = & \kappa^\ast_0 - \frac{1}{\xi}\frac{\pi c}{3 L^2}
+ O(L^{-3})
\label{gsem} 																																																																											
\end{eqnarray}
with the constant
\begin{equation}
\kappa^\ast_0 = \frac{\nu}{2}  - 1
+ \frac{1}{\pi}
 \left(\frac{\arcsin{\nu}}{\nu}+ \sqrt{1-\nu^2}\right),
\end{equation}
the sound velocity
\begin{equation}
\xi = \frac{1}{\sqrt{1-\nu^{2}}},
\label{svdef}
\end{equation}
and the central charge $c=1/2$ characterizing the Ising quantum chain 
in a transverse field.
\end{prop}

This is essentially a well-established classical result \cite{Henk87}. A somewhat simpler proof is provided below.\\

\Proof
With the parameter $\nu$ \eqref{nudef} and the binomial series expansion
\begin{equation}
\sqrt{1-x^2} = \sum_{m=0}^\infty (-1)^m {1/2 \choose m} x^{2m}
\end{equation}
it follows that
\begin{eqnarray*}
K_{L}(s_c) & = & \frac{\nu}{2}  - 1 + \sum_{r=0}^{L/2-1} \lambda^{+(0)}_{L,r} \\
& = & \frac{\nu}{2}  - 1 
+ \frac{1}{2} \sum_{m=0}^\infty (-1)^m \nu^{2m} {1/2 \choose m}  \Phi_m(L/2)
\end{eqnarray*}
and by Lemma \ref{Lem:BmN}
\begin{eqnarray*}
K_{L}(s_c) 
& = & \frac{\nu}{2}  - 1
+ \frac{1}{2} \sum_{m=0}^\infty (-1)^m \nu^{2m} {1/2 \choose m}   \left(\frac{4}{\pi (2m+1)}  +  \frac{\pi }{6 L^2}\right) \\
& = & \frac{\nu}{2}  - 1
+ \frac{2}{\pi}\sum_{m=0}^\infty (-1)^m  {1/2 \choose m}   \frac{\nu^{2m}}{2m+1} +  \sqrt{1-\nu^2} \frac{\pi }{12 L^2} \\
& = & \frac{\nu}{2}  - 1
+ \frac{2}{\pi \nu}\sum_{m=0}^\infty (-1)^m  {1/2 \choose m}  
\int_{0}^{\nu} x^{2m} \, \rmd x + \sqrt{1-\nu^2} \frac{\pi}{12 L^2} \\
& = & \frac{\nu}{2}  - 1
+ \frac{2}{\pi \nu}
\int_{0}^{\nu} \sqrt{1-x^{2}} \, \rmd x + \sqrt{1-\nu^2} \frac{\pi}{12 L^2} 
\end{eqnarray*}
The assertion \eqref{gsep} follows by using
\be
\int_0^{u} \rmd y \sqrt{1-y^2}  = \half (\arcsin{u}+ u \sqrt{1-u^2})
\end{equation}
and observing that the limits $\nu\to 1$ and $\nu\to 0$  
are well-defined. 
\qed

\subsection{Critical cumulants}

Combining the two Lemmas \ref{Lemma:FdB1} and \ref{Lemma:FdB2} 
yields exact expressions for the cumulants for any value of $s$. Here we are interested in the evaluation of the cumulants at the 
critical point.
With the the coefficients 
\begin{equation}
\lambda^{(n)}_{L,r} := \frac{d^n}{ds^n} \left.\Lambda_{L,r}(s)\right\vert_{s=s_c} 
\label{dcdrdef}
\end{equation}
the expansion of the cumulant generating function \eqref{gse}
around the critical point $s_c=\nu$ reads
\begin{equation}
K^{c}_{L}(v) = \frac{2}{L} \sum_{r=0}^{L/2-1} \sum_{n=1}^\infty \lambda^{(n)}_{L,r} \frac{v^n}{n!}
\end{equation}
and \eqref{kpmnLs} becomes
\begin{equation}
\kappa^{c}_{n}(L) = \left. \frac{d^n}{dv^n} K^c(L,v)\right\vert_{v=0}
= \frac{2}{L} \sum_{r=0}^{L/2-1} \lambda^{(n)}_{L,r}
\label{knL}
\end{equation}
for all $n\geq 1$. 

\subsubsection{Exact expressions for the critical point}

\begin{prop}
\label{Prop:dcdr}
The derivatives \eqref{dcdrdef} of the critical dispersion relation
 are given by
\begin{eqnarray}
\lambda^{(n)}_{L,r}
& = & \sum_{m=0}^{n-1} 
 \frac{\nu^{n-m}}{2^{2n-2m-1}} 
S(n,n-m) \nonumber \\
& & \times
\sum_{k= \lfloor \frac{n+m+1}{2} \rfloor}^{n}
\frac{(-1)^{k-m+1} (n-m)! (2k-2m-2)!}{(2k-n-m)! (n-k)!(k-m-1)!} \nonumber \\
& & \times (\mathsf{s}_{L,r})^{2k+1-2n}
\frac{ \left(1-2\nu (\mathsf{c}_{L,r})^2 \right)^{2k-n-m}  \left(1-2\mathsf{c}_{L,r} \right)^{2n-2k}}{\left(1 - \nu^2 (\mathsf{c}_{L,r})^{2}\right)^{k-m-1/2}}  
\label{dcdr}
\end{eqnarray}
for all $n\geq 1$.
\end{prop}

\Proof
With the quadratic function 
\begin{equation}
f_{L,r}(x) := \nu^2 + 2 (1-\nu) (1 -\mathsf{c}_{L/2,r}) 
+ 2x (1-\nu - \mathsf{c}_{L/2,r}) + x^2 (\mathsf{c}_{L/2,r})^2 
\end{equation}
of the variable $x$
the dispersion relation \eqref{betahr} is given by
\begin{equation}
\Lambda_{L,r}(s) = \frac{1}{2} \sqrt{f_{L,r}(\rme^s)} .
\end{equation}
By Lemma \ref{Lemma:FdB1}
\begin{equation}
\frac{d^n}{ds^n} \Lambda_{L,r}(s) = \sum_{m=1}^{n} S(n,m) x^{m} \left.\frac{d^m}{dx^m} \sqrt{f_{L,r}(x)}\right\vert_{x=\rme^{s}} 
\end{equation}
where Lemma \ref{Lemma:FdB2} yields for $m\geq 1$
\begin{eqnarray}
\frac{d^m}{dx^m} \sqrt{f_{L,r}(x)} 
& = & \sum_{k= \lfloor \frac{m+1}{2} \rfloor}^m
\frac{(-1)^{k+1}}{2^{m-1}} 
\frac{m!(2k-2)!}{(2k-m)! (m-k)!(k-1)!} \nonumber \\
& & \times
\frac{\left(1-\nu - \mathsf{c}_{L/2,r} + x (\mathsf{c}_{L/2,r})^2\right)^{2k-m}(\mathsf{c}_{L/2,r})^{2m-2k}
}{\left(f_{L,r}(x)\right)^{k-1/2}} .
\end{eqnarray}

At the critical point $x=\nu$ one has
\begin{eqnarray}
f_{L,r}(\nu) & = &  4 (\mathsf{s}_{L,r})^2 \left(1 - \nu^2 (\mathsf{c}_{L,r})^2\right) \\
f_{L,r}{\hspace*{-4pt}}'(\nu) & = & 4 (\mathsf{s}_{L,r})^2 \left(1-2\nu (\mathsf{c}_{L,r})^2\right) \\
f_{L,r}{\hspace*{-4pt}}''(\nu) & = & 2 \left(1-2(\mathsf{c}_{L,r})^2\right)^2
\end{eqnarray}
where the trigonometric identities $\mathsf{s}_{L/2,r} = 2 \mathsf{s}_{L,r}\mathsf{c}_{L,r}$ and $\mathsf{c}_{L/2,r} = 2 (\mathsf{c}_{L,r})^2-1=1 - 2(\mathsf{s}_{L,r})^2$
are used. 
\qed

\begin{rem}
As seen below, only the Stirling numbers of the second kind 
\begin{eqnarray}
S(n,n) & = & 1 \\
S(n,n-1) & = & {n \choose 2} 
\end{eqnarray}  
are relevant for the asymptotic behaviour of the cumulants.
\end{rem}

\subsubsection{Asymptotic behaviour}

\paragraph{(i) Critical mean activity}

For any $s\in\R$
the mean activity $A_{L}(s)$ is the first cumulant, i.e.,
the first derivative 
\begin{equation}
A_{L}(s)  = \frac{d}{ds} K_{L}(s)  
\label{Alsdef}
\end{equation}
of the cumulant generating function w.r.t. $s$.
Straightforward computation yields
\begin{equation}
A_{L}(s) = 
 \frac{\rme^s}{L} \sum_{r=0}^{L/2-1} 
\frac{ \left( 1 - (\rme^{s} + \tilde{\mu}) \mathsf{c}_{L/2,r}\right)
\mathsf{c}_{L/2,r}
 -  \tilde{\mu} (\mathsf{s}_{L/2,r})^2}{
\sqrt{
\left( 1 - (\rme^{s} +\tilde{\mu}) \mathsf{c}_{L/2,r}\right)^2 
+  (2 \rme^{s} +\tilde{\mu}) \tilde{\mu} (\mathsf{s}_{L/2,r})^2}}.
\label{ALs}
\end{equation}
At the critical point we have $A^{c}(L) :=A_{L}(s_c) = \kappa_{1}^{c}(L)$, i.e.,
\begin{equation}
A^{c}(L)  = \frac{2}{L} \sum_{r=0}^{L/2-1} \lambda_{L,r}^{\pm(1)}.
\end{equation}

\begin{prop}
\label{Prop:mca}
To leading and subleading order in $1/L$ the critical mean activity 
in the range $0 \leq \nu <1$ is given by
\begin{eqnarray}
A^{c}(L) & = & \kappa^\ast_1
+ \frac{\nu \left(1-2 \nu \right) }{\sqrt{1-\nu^2}}
\frac{\pi}{24 L^2}
\label{mcap}
\end{eqnarray}
with the asymptotic critical mean activity 
\begin{equation}
\kappa^\ast_1 = \frac{1}{\pi} 
\left(\frac{1}{\sqrt{1-\nu^2}} - \frac{1-\nu}{\nu}
\arcsin{\nu} \right) .
\label{k1ast}
\end{equation}
\end{prop}

\Proof By Proposition  \ref{Prop:dcdr}
\begin{equation}
A^{c}(L)  =
\frac{\nu}{L} \sum_{r=0}^{L/2-1}  \mathsf{s}_{L,r}
\frac{ 1-2\nu (\mathsf{c}_{L,r})^2}{\left( 1 - \nu^2 (\mathsf{c}_{L,r})^2 \right)^{1/2}} .
\label{dcdr1}
\end{equation}
With the Taylor expansion
\begin{equation}
\frac{1}{\sqrt{1-x^2}} = \sum_{n=0}^\infty (-1)^n {-1/2 \choose n} x^{2n}
\end{equation}
one gets
\begin{eqnarray*}
A^{c}(L) 
& = & \frac{1}{2} \sum_{n=0}^\infty (-1)^n {-1/2 \choose n} \nu^{2n+1}
\left(\Phi_n(L/2)-2 \nu \Phi_{n+1}(L/2)\right)
\nonumber \\
& = & \frac{1}{\pi} \sum_{n=0}^\infty (-1)^n {-1/2 \choose n} 
\left(\frac{\nu^{2n+1}}{2n+1} - \frac{2 \nu^{2n+2}}{2n+3}\right)
\nonumber \\
& & + \frac{1}{2} \nu \left(1-2 \nu \right) \sum_{n=0}^\infty (-1)^n {-1/2 \choose n} \nu^{2n}
\frac{\pi}{12 L^2}
\nonumber \\
& = & \frac{1}{\pi} \sum_{n=0}^\infty (-1)^n {-1/2 \choose n} 
\left(\int_{0}^{\nu} x^{2n} \, \rmd x - \frac{2 }{\nu}
\int_{0}^{\nu} x^{2n+2} \, \rmd x\right)
\nonumber \\
& & + \frac{\nu \left(1-2 \nu \right) }{2 \sqrt{1-\nu^2}}
\frac{\pi}{12 L^2}
\nonumber \\
& = & \frac{1}{\pi} 
\left(\int_{0}^{\nu} \frac{1}{\sqrt{1-x^2}} \, \rmd x - \frac{2 }{\nu}
\int_{0}^{\nu} \frac{x^2}{\sqrt{1-x^2}} \, \rmd x\right)
\nonumber \\
& & + \frac{\nu \left(1-2 \nu \right) }{2 \sqrt{1-\nu^2}}
\frac{\pi}{12 L^2} .
\end{eqnarray*}
The assertion \eqref{mcap} then follows from the indefinite integrals
\begin{eqnarray*}
\int \frac{1}{\sqrt{1-x^2}} \, \rmd x & = & \arcsin{x} \\
\int \frac{x^2}{\sqrt{1-x^2}} \, \rmd x & = & \frac{1}{2}\arcsin{x} - \frac{1}{2}  \frac{x}{\sqrt{1-x^2}}
\end{eqnarray*}
and observing that the limit $ \nu \to 0$ is well-defined.
\qed


\paragraph{(ii) Critical Variance}
The variance $V^{c}(L) $ of the activity at the critical point is the second cumulant, i.e.,
\begin{equation}
V^{c}(L) = \frac{2}{L} \sum_{r=0}^{L/2-1} \lambda_{L,r}^{(2)}
\end{equation}
where the constants $\lambda_{L,r}^{(2)}$ are given by \eqref{dcdr}.

\begin{prop}
\label{Prop:Var}
For $\nu\in[0,1)$ the variance of the activity at the critical point
is logarithmically divergent in $L$ and given by
\begin{equation}
V^{c}(L) \sim
\frac{ \nu^{2}}{2 \left(1 - \nu^2 \right)^{1/2} \pi} \ln{L}.
\label{Var}
\end{equation}
 to leading order in $L$.
\end{prop}

\Proof 
According to Proposition  \ref{Prop:dcdr} one can write $\lambda_{L,r}^{(2)} = 
\lambda_{L,r,1}^{(2)} + \lambda_{L,r,2}^{(2)}$ where
\begin{eqnarray}
\lambda_{L,r,1}^{(2)}
& = &  \frac{1}{\mathsf{s}_{L,r}} 
\frac{  \left(1-2\mathsf{c}_{L,r} \right)^{2}}{\left(1 - \nu^2 (\mathsf{c}_{L,r})^2\right)^{1/2}}  \frac{\nu^{2}}{4} 
\label{dcdr21} \\
\lambda_{L,r,2}^{(2)}
& = & 
\mathsf{s}_{L,r}
\frac{1-2\nu (\mathsf{c}_{L,r})^2}{\left(1 - \nu^2 (\mathsf{c}_{L,r})^2\right)^{3/2}}  \frac{\nu (2-\nu)}{4} 
\label{dcdr22}
\end{eqnarray}
The second term $\lambda_{L,r,2}^{(2)}$ is of a form similar to the 
$\lambda_{L,r}^{(2)}$ appearing in the expression \eqref{dcdr1} for the 
mean activity and yields by
Lemma \ref{Lem:BmN} a subleading contribution of order 1 to the variance.
The key point in estimating the first term $\lambda_{L,r,1}^{(2)}$ is a saddle-point argument: Because of the inverse power of the sine function
$\mathsf{s}_{L,r}$ in $\lambda_{L,r,1}^{(2)}$,
the dominant contribution to the sum \eqref{knL} comes from small 
arguments of the sine, i.e., from the integers $r$ that are small 
compared to $L$. 
The second observation is that for fixed $r$ the expansion of the cosine
$\mathsf{c}_{L,r} = 1 - r^2 O(L^{-2})$ so that the subleading terms can be neglected. On the other hand, for large values of $r$, the factor
containing the cosines is uniformly bounded, while the sine is of subleading order. This generates at most $L$ subleading
terms which can be neglected. Consequently,  the terms
\begin{eqnarray}
\lambda_{L,r}^{(2)}
& \sim & \frac{L}{2\pi(2r+1}
\frac{ \nu^{2}}{\left(1 - \nu^2 \right)^{1/2}} 
\end{eqnarray}
with arbitrary but fixed values of the summation index $r$ yield the
 leading order contribution to the variance obtained from the leading
behavior
\begin{equation}
\frac{2}{L}
\sum_{r=1}^{L/2-1}
\frac{L}{\pi(2r+1)}
\sim 
 \frac{1}{\pi} \ln{L} 
\end{equation}
of the sum over $r$. 
\qed

\paragraph{(iii) Critical skewness}

By definition, the skewness 
\begin{equation}
S^{c}(L) = \frac{\kappa^{c}_{3}(L)}{(\kappa^{c}_{2}(L))^{3/2}}
\label{skewdef}
\end{equation}
of the activity distribution at the critical point
is given by the third cumulant and the variance. To compute
the large-$L$ behaviour of the critical skewness 
 we  use Proposition \ref{Prop:Var} and compute
the third cumulant.

\begin{prop}
\label{Prop:skew}
For $\nu\in(0,1)$ the critical skewness vanishes as $L\to\infty$
and is given by
\begin{equation}
S^{c}(L) \sim
 \frac{3\left(2-\nu\right)}{\nu \left(1 - \nu^2\right)^{3/4}} 
\sqrt{\frac{\pi}{2\ln{L}}}
\label{skew}
\end{equation}
 to leading order in $L$.
\end{prop}

\Proof To arrive at \eqref{skew} the leading behaviour of the third
cumulant 
\begin{equation}
\kappa^{c}_{3}(L) = \frac{2}{L} \sum_{r=0}^{L/2-1} \lambda_{L,r}^{(3)}
\end{equation}
needs to be determined. This follows entirely along the lines
of the proof of the variance. The leading term in $\lambda_{L,r}^{(3)}$ comes from $m=0$ and $m=1$ with $k=2$ in \eqref{dcdr} which
is the term proportional to $\mathsf{s}_r^{-1}(L)$. As in the
case of the variance, this leads to a logarithmic divergence
of $\kappa^{c}_{3}(L)$
in $L$. Straighforward computation of the coefficient, given
by taking the cosine function at value 1, yields
\begin{equation}
\kappa^{c}_{3}(L) \sim \frac{3}{4}
 \frac{\nu^{2} \left(2-\nu\right)}{\left(1 - \nu^2\right)^{3/2} \pi}
\ln{L}
\end{equation}
to leading order in $L$ which together with the definition 
\eqref{skewdef} and the variance \eqref{Var} yields \eqref{skew}.
\qed

\paragraph{(iv) General cumulants at the critical point}

With the arguments used in the proofs of Propositions \ref{Prop:mca} - 
\ref{Prop:skew} and the exact result \eqref{dcdr} of Proposition 
\ref{Prop:dcdr} we are now in a position to state and prove the main
result of this Section.

\begin{theo}
\label{Theo:ucn}
For $\nu\in[0,1)$
the cumulants at the critical point are given 
to leading order in $L$  by
\begin{eqnarray}
\kappa^{c}_{1}(L) & \sim & \kappa^\ast_{1} \\
\kappa^{c}_{2}(L) & \sim & \kappa^\ast_{2}  \ln{L}\\
\kappa^{c}_{3}(L) & \sim & \kappa^\ast_{3} \ln{L} \\
\kappa^{c}_{2n}(L) & \sim & \kappa^{\ast}_{2n}   L^{2n-2}, \quad
n>2 \label{kceven} \\
\kappa^{c}_{2n+1}(L) & \sim & \kappa^{\ast}_{2n+1}  L^{2n-2}, \quad
n>2 
\label{kcodd}
\end{eqnarray}
with the mean activity $\kappa^\ast_1$ \eqref{k1ast} and the rescaled
cumulants
\begin{eqnarray}
\kappa^{\ast}_{2n}  & = & (2n)! \alpha_n \beta_{2n}(\nu)
\label{kasteven} \\
\kappa^{\ast}_{2n+1}  & = & (2n+1)! \alpha_n \beta_{2n+1}(\nu) 
\label{kastodd}
\end{eqnarray}
given for $n>0$ by 
\begin{eqnarray}
\alpha_1 & := & \frac{1}{2\pi}
\label{a1def} \\
\alpha_n & := & \frac{1-2^{1-2n}}{\pi^{2n-1}}  \zeta(2n-1), \quad n>1
\label{anpdef} \\
\beta_{2n} & := & \frac{\nu^{2n}  \binom{1/2}{n}}{2 \left(1 - \nu^2 \right)^{n-1/2}}
\label{bedef} \\
\beta_{2n+1} & := &  \frac{\nu^{2n} \left[\nu\left(1-2\nu\right)  {-1/2 \choose n}
+ \left(1 - \nu^2\right) {-1/2 \choose n-1}\right] }{4\left(1 - \nu^2\right)^{n+1/2}}
\label{bodef}
\end{eqnarray}
and the Riemann $\zeta$-function $\zeta(\cdot)$.
\end{theo}

\Proof For $n=1$ the assertion is an immediate consequence of Proposition 
\ref{Prop:mca}. The cases $n=2$ and $n=3$ are treated in Proposition
\ref{Prop:Var} and in the proof of Proposition \ref{Prop:skew}. 
For all higher cumulants the proof is analogous to the proof of the variance.
The dominant contribution to the sum \eqref{knL} comes from the term in \eqref{dcdr} with the 
largest negative power of the sine-function, which corresponds for even
cumulants for to $m=0$ and $k=n$ in the double sum \eqref{dcdr}
which is
\begin{equation}
\lambda_{L,r}^{(2n)} \sim 
\frac{(-1)^{n+1} (2n)! (2n-2)! \nu^{2n} \left(1-2\mathsf{c}_{L,r} \right)^{2n}}{2^{4n-1}n!(n-1)!\left(1 - \nu^2(\mathsf{c}_{L,r})^2 \right)^{n-1/2}} 
(\mathsf{s}_{L,r})^{1-2n}
\end{equation}
with arbitrary but fixed values of the summation index $r$. 
Expanding $\mathsf{s}_r^{1-2n}(L)$ and the cosine functions to leading order
one arrives at
\begin{eqnarray*}
\lambda_{L,r}^{+(2n)} &  \sim  &
\frac{(-1)^{n+1} (2n)! (2n-2)! \nu^{2n} }{2^{4n-1}n!(n-1)!\left(1 - \nu^2 \right)^{n-1/2}} 
\left(\frac{2L}{\pi(2r+1)}\right)^{2n-1} \\
\lambda_{L,r}^{-(2n)} &  \sim  &
\frac{(-1)^{n+1} (2n)! (2n-2)! \nu^{2n} }{2^{4n-1}n!(n-1)!\left(1 - \nu^2 \right)^{n-1/2}} 
\left(\frac{L}{\pi r}\right)^{2n-1}  
\end{eqnarray*}
One has for $n\geq 2$
\begin{eqnarray*}
\frac{2}{L}
\sum_{r=0}^{L/2-1}
\left(\frac{2L}{\pi(2r+1)}\right)^{2n-1}
& \sim & \frac{4 (2L)^{2n-2} (1-2^{1-2n}) \zeta(2n-1) }{\pi^{2n-1}} \\
\frac{2}{L}
\sum_{r=0}^{L/2-1}
\left(\frac{L}{\pi r}\right)^{2n-1}
& \sim & \frac{4 (2L)^{2n-2} 2^{1-2n} \zeta(2n-1) }{\pi^{2n-1}} \\
\end{eqnarray*}
The identity
\begin{equation}
\binom{1/2}{n} = \frac{(-1)^{n-1}}{2^{2n-1} n} {2n-2 \choose n-1}, \quad n\geq 1
\end{equation}
then yields the general formula \eqref{kceven} with \eqref{kasteven}
for $n \geq 1$.

Similar considerations show that for odd cumulants only the terms
with $m=0$ and $m=1$ and $k=n+1$ yield the leading behaviour.
Therefore with the  identity
\begin{equation}
\binom{-1/2}{n} = \frac{(-1)^{n}}{2^{2n}} {2n \choose n}
\end{equation}
one finds
\begin{eqnarray*}
\lambda_{L,r}^{(2n+1)}
& \sim & (\mathsf{s}_{L,r})^{1-2n} \frac{(2n+1)! }{2^{2n}}
 \frac{\nu^{2n+1}}{\left(1 - \nu^2\right)^{n+1/2}} 
 \nonumber \\
& & \times \nu \left(1-2\nu\right) 
\frac{(-1)^{n}(2n)!}{2^{2n} n!n!}  \nonumber \\[2mm]
&  & - (\mathsf{s}_{L,r})^{1-2n}  \frac{(2n+1)! }{2^{2n+1}}
 \frac{\nu^{2n}}{\left(1 - \nu^2\right)^{n+1/2}} 
\nonumber \\
& & \times \left(1 - \nu^2\right) 
\frac{(-1)^{n} (2n-2)!}{2^{2n-2}(n-1)!(n-1)!} \nonumber \\
& \sim & \left(\frac{2L}{\pi(2r+(1\pm 1)/2)}\right)^{2n-1} \frac{(2n+1)! }{2^{2n+1}}
 \frac{\nu^{2n}}{\left(1 - \nu^2\right)^{n+1/2}} 
 \nonumber \\
& & \times \left(\nu\left(1-2\nu\right)  {-1/2 \choose n}
+ \left(1 - \nu^2\right) {-1/2 \choose n-1}\right) 
\end{eqnarray*}
which yields  \eqref{kcodd} with
 \eqref{kastodd} for $n\geq 1$.
\qed

\begin{cor} For $k\geq 1$ the cumulants $\kappa^{c}_{2k}$ and $\kappa^{c}_{2k+1}$ are of equal divergent order in $L$ so that 
for $n\geq 3$
\begin{equation}
\lim_{L\to\infty} L^{2-n} \kappa^{c}_{L,n} =
\left\{ \ba{ll}
0 & n \mbox{ odd} \\
\kappa^{\ast}_{n} & n = \mbox{ even} .
\ea\right.
\end{equation}
\end{cor}

\begin{rem}
With the convention ${-1/2\choose -1}=0$ the coefficients $\beta_0= \sqrt{1-\nu^2}/2$
and $\beta_1=\nu^3(1-2\nu)/(4\sqrt{1-\nu^2})$ are well-defined.
\end{rem}

\subsection{Universal cumulant generating function}

To establish universality of the activity distribution near the critical
point consider the
function
\begin{equation}
h(u) := \sum_{r=0}^\infty \left(\sqrt{u^2 + \pi^2 (2r+1)^2}   -  \pi(2r+1) -  \frac{ u^{2} }{ 2 \pi(2r+1)}  \right)
\end{equation}
which is well-defined through the Taylor expansion of the square root
in $u^2$.

\begin{theo}
\label{Theo:ucgf}
The function
\begin{equation}
K_0(L,u) := \xi \left[K(L,s_c+u/L) - \kappa^\ast_0 -  \frac{u}{L} \kappa^c_1(L) - \frac{u^2}{2L^2} \kappa^c_2(L)\right]
\end{equation}
with the sound velocity $\xi=1/ \sqrt{1-\nu^2}$ has the 
universal scaling limit
\begin{eqnarray}
\lim_{L\to\infty} L^2 K_0(L,u) & = & \frac{\pi c}{6} + h(\sqrt{\xi^2-1} \, u)
\label{ucgf}
\end{eqnarray}
with the central charge $c=1/2$ and the parameter free scaling function $h(\cdot)$.
\end{theo}

\Proof 
By Proposition \ref{Prop:dcdr}
\begin{eqnarray*}
K^c(L,u/L) & = & \frac{u}{L} \kappa^c_1(L) + \frac{u^2}{2L^2} \kappa^c_2(L)
+  \frac{u^3}{6L^3} \kappa^c_3(L) \\
& & + \sum_{n=2}^\infty  \frac{2}{L} \sum_{r=0}^{L/2-1} 
\frac{\lambda_r^{(2n)}(L)}{L^{2n}} \frac{u^{2n}}{(2n)!} \\
& & + \frac{1}{L}\sum_{n=2}^\infty  \frac{2}{L} \sum_{r=0}^{L/2-1} 
\frac{\lambda_r^{(2n+1)}(L)}{L^{2n}} \frac{u^{2n+1}}{(2n+1)!}
\end{eqnarray*}
Therefore
\begin{eqnarray*}
\tilde{K}_0(L,u) & := & K(L,s_c+u/L) - K_{L}(s_c) -  \frac{u}{L} \kappa^c_1(L) - \frac{u^2}{2L^2} \kappa^c_2(L) \\
& = &  \sum_{n=2}^\infty  \frac{2}{L} \sum_{r=0}^{L/2-1} 
\frac{\lambda_r^{(2n)}(L)}{L^{2n}} \frac{u^{2n}}{(2n)!} \\
& & + \frac{1}{L}\sum_{n=2}^\infty  \frac{2}{L} \sum_{r=0}^{L/2-1} 
\frac{\lambda_r^{(2n+1)}(L)}{L^{2n}} \frac{u^{2n+1}}{(2n+1)!}
\end{eqnarray*}
and by Theorem \ref{Theo:ucn}
\begin{eqnarray}
\lim_{L\to\infty} L^2 \tilde{K}_0(L,s_c+u/L) & = & \sum_{n=2}^\infty \kappa^\ast_{2n} \frac{u^{2n}}{(2n)!} 
\end{eqnarray}
On the other hand, by Taylor expansion one gets
\begin{equation}
2w \sqrt{(1-\nu^2)} h(\vartheta u) = 
 \sum_{n=2}^\infty  \kappa^\ast_{2n}(\nu) \frac{u^{2n}}{(2n)!}  .
\end{equation}
The equality \eqref{ucgf} then follows from the leading-order finite-size corrections \eqref{gsep} and \eqref{mcap}.
\qed

\begin{figure}[h]
    \centering 
    \includegraphics[width=0.7\textwidth]{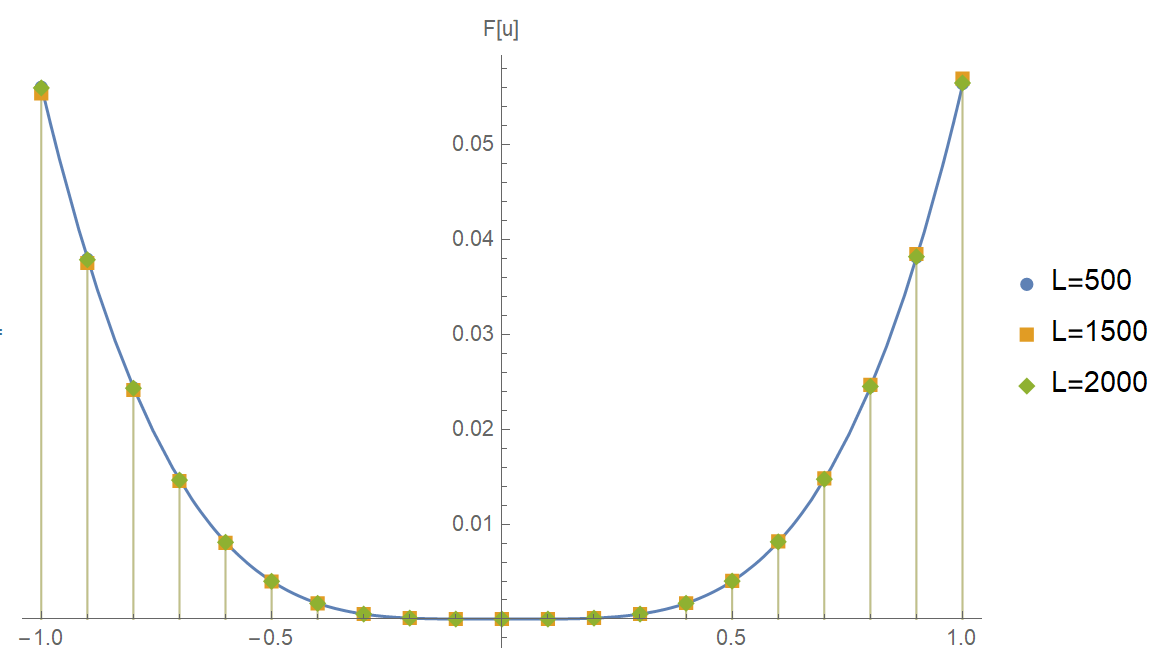}
    \caption{Blue curve: The scaling function $h(\sqrt{\xi^2-1} \, u)$ for $\nu=0.9$. The values at the vertical lines  are numerical evaluations of the
exact generating function for different L. They all collapse onto the same
point along the blue curve.}
    \label{fig:example1} 
\end{figure}

The theorem establishes a direct link between the cumulant generating function
{\it near} the critical point and universal quantities
{\it at the critical point}, obtained by substracting the
leading non-universal contributions up to order two from 
the generating function for the cumulants of order 4 and higher. It can be recast as 
\begin{eqnarray}
\lim_{L\to\infty} L^2 \tilde{K}_0(L,u) & = &  h(\sqrt{\xi^2-1} \, u).
\label{ucgf2}
\end{eqnarray}
Numerically, already moderately high system sizes reproduce the
limiting behaviour, as shown in Fig.~\ref{fig:example1}.

\begin{rem}
Consider the
coefficients 
\begin{eqnarray}
\check{g}_r(u) & := & w u \frac{ \nu^2u^2 + \nu (1-2\nu) \pi^2 (2r+1)^2 }
{ \sqrt{\nu^2u^2 + (1-\nu^2) \pi^2 (2r+1)^2}}
\end{eqnarray}
and the function
\begin{eqnarray}
g(u) & := & \sum_{r=0}^\infty \left( \check{g}_r(u) -  2\beta_{1} \pi(2r+1)  u
-  \beta_{3} \frac{2}{\pi(2r+1)}  u^{3} \right)
\end{eqnarray}
By Taylor expansion of the square root one finds that
\begin{eqnarray*}
g(u) & =&  \sum_{n=2}^\infty  \kappa^\ast_{2n+1}(\nu) \frac{u^{2n+1}}{(2n+1)!} .
\end{eqnarray*}
This generating function is non-universal since it depends on the microscopic
parameter $\nu$.
\end{rem}

\section*{Acknowledgements}
G.M.S. thanks the LPCT at the Universit\'e de Lorraine where part of this work was done, for kind 
hospitality. This work was supported by FCT (Portugal) 
through project UIDB/04459/2020, 
doi 10-54499/UIDP/04459/2020
and by the FCT Grants 2020.03953.CEECIND and 
2022.09232.PTDC.

\end{document}